# Transforming Single Domain Magnetic $CoFe_2O_4$ Nanoparticles from Hydrophobic to Hydrophilic By Novel Mechanochemical Ligand Exchange

Sandeep Munjal and Neeraj Khare*

Department of Physics, Indian Institute of Technology Delhi, Hauz Khas, New Delhi-110016, India.

**Abstract**

Single phase, uniform size (~9 nm) Cobalt Ferrite (CFO) nanoparticles have been synthesized by hydrothermal synthesis using oleic acid as a surfactant. The as synthesized oleic acid coated CFO (OA-CFO) nanoparticles were well dispersible in nonpolar solvents but not dispersible in water. The OA-CFO nanoparticles have been successfully transformed to highly water dispersible citric acid coated CFO (CA-CFO) nanoparticles using a novel single step ligand exchange process by mechanochemical milling, in which small chain citric acid molecules replace the original large chain oleic acid molecules available on CFO nanoparticles. The OA-CFO nanoparticle's hexane solution and CA-CFO nanoparticle's water solution remain stable even after six months, and shows no agglomeration and their dispersion stability was confirmed by zeta potential measurements. The contact angle measurement shows that OA-CFO nanoparticles are hydrophobic whereas CA-CFO nanoparticles are superhydrophilic in nature. The potentiality of as synthesized OA-CFO and mechanochemically transformed CA-CFO nanoparticles for the demulsification of highly stabilized water-in-oil and oil-in-water emulsions has been demonstrated.

*Authors to whom all correspondence should be addressed.

E-mail: nkhare@physics.iitd.ernet.in

## 1. Introduction

Magnetic nanoparticles have attracted a great attention of scientific community due to their potentiality in many advanced and novel applications such as for high-density data storage (Reiss and Hütten 2005), targeted drug delivery (Arruebo et al. 2007), catalysis (Kharisov 2014), as contrast enhancement agents in magnetic resonance imaging (Wang et al. 2011), in magnetic hyperthermia treatments as heat mediators (Brollo et al. 2016), for resistive switching applications (Munjal, Kumari, et al. 2016) ,and for oil−water multiphase separation etc. The surface of these magnetic nanoparticles is functionalized by organic (Wang et al. 2011) or inorganic shell as per the requirement of the application, and by changing the surface chemistry the properties like mutual interaction between the nanoparticles or wettability behaviour can be influenced (Bajwa et al. 2016). In order to control the size/shape of these nanoparticles surfactants are used during the growth process. These surfactant molecules acts as a stabilizing agent that prevent the agglomeration and slow down the growth rate of the nanoparticles (Kumar et al. 2015). Such nanoparticles, synthesized in the presence of surfactant are dispersible in organic /nonpolar solvents like hexane, toluene etc. On the other hand, this hydrophobic nature of the magnetic nanoparticles limits their use in many applications including the application in biochemistry, biomedical (Huang and Juang 2011) and thin film fabrication. For all these applications the nanoparticles must be dispersible in polar solvents like water or ethanol and this requires a further modification of the nanoparticles suface in order to make them hydrophilic. This modification of surface can be done in two ways; (a) binding an amphiphilic molecule to the original surfactant layer through hydrophobic interactions and forming a micellar structure that encapsulates the magnetic nanoparticles or (b) replacing the native hydrophobic/Oleophilic surfactants a by small chain hydrophilic molecule that have higher affinity for the metal ion present in the magnetic nanoparticles (Kumar et al. 2015). The later one is called ligand exchange process. The ligand, replacing the native hydrophobic surfactant must have an anchoring group (phosphonic acid, dopamine etc.) that binds to the surface of magnetic nanoparticles and the other end of this ligand must be hydrophilic that is exposed to the surrounding $H_2O$ molecules and gives a colloidal stability to the magnetic nanoparticles in the aqueous medium. However, the conventional methods of ligand exchange may take many hours and are generally multistep (Hatakeyama et al. 2011). The development of simple methods to generate magnetic nanoparticles stable in aqueous environments remains the subject of vigorous inquiry.

Iron oxide magnetic nanoparticles such as $\mathrm{Fe_3O_4}$ (Tamer et al. 2010) and $\gamma-\mathrm{Fe_2O_3}$ (Hergt et al. 2004) have been extensively explored for these applications. The another alternate can be spinel ferrite [ $MFe_2O_4$, M=Co, Ni, Zn etc] magnetic nanoparticles, and specially cobalt ferrite ($CoFe_2O_4$) due to its large Curie temperature, high effective anisotropy and moderate saturation magnetization (Bricen et al. 2012) . $CoFe_2O_4$ (CFO) has an inverse spinel structure with general formula $AB_2O_4$ (A = Fe and B = Co, Fe) where half of the $Fe^{3+}$ occupies the octahedral sites and the other half $Fe^{3+}$ occupies the tetrahedral sites whereas all the $Co^{2+}$ occupies the octahedral sites (Munjal, Khare, et al. 2016). Several techniques such as microemulsion (Mathew and Juang 2007), coprecipitation (Kim, Kim, and Lee 2003) , ball milling (Manova et al. 2004) , sol−gel (Lavela and Tirado 2007) , thermal decomposition (Kalpanadevi, Sinduja, and Manimekalai 2014) and sonochemical (Saffari et al. 2014) method have been employed for the synthesis of magnetic nanoparticles but all these synthesis methods often produce larger size nanoparticles with wide particle size distribution. But most of the above stated applications requires a narrow particle size distribution, as the performance of the magnetic nanoparticles strongly depends upon the particle size and particle size distribution.

In the present work, we report the synthesis of uniform size oleic acid coated CFO (OA-CFO) nanoparticles by hydrothermal synthesis method (Chaudhary, Khare, and Vankar 2016), which are not water soluble and the conversion of these OA-CFO nanoparticles through a rapid and novel mechanochemical ligand exchange process using citric acid as a hydrophilic ligand to replace oleic acid from the surface of OA-CFO nanoparticles in a single step. Due to the short carbon chain and the presence of multiple carboxylic groups the citric acid becomes a suitable candidate for the ligand exchange process. A direct evidence of wettability of these $CoFe_2O_4$ magnetic nanoparticles has been given by contact angle measurements. The use of OA-CFO and CA-CFO nanoparticles for the demulsification of highly stabilized oil in water emulsions and water in oil emulsions is also demonstrated.

Conventional ligand exchange methods like stirring based method (Munjal, Khare, et al. 2016) or Solid-state photochemical ligand exchange (Loim, N.M., Khruscheva, N.S., Lukashov 1999) suffers from time consumption and may take from 24 h to 48 h. Our presented mechanochemical ligand exchange process is rapid and takes only ~ 30 minutes. We have repeated this method several times and each time the results are reproducible. Application of our process may have impacts on the modification of hydrophobic magnetic nanoparticles and preparation of hydrophilic magnetic nanoparticles.

The synthesized CA-CFO has many practical applications in biomedical. These water dispersible magnetic nanoparticles can be used (a) for targeted drug delivery (b) as contrast enhancement agents in magnetic resonance imaging (MRI) and (c) in hyperthermia treatments as heat mediators. Beside all these applications, these nanoparticles can also be used as Draw Solutes in Forward Osmosis for Water reuse (Ge et al. 2011).

## 2. Experimental Section

### 2.1 Synthesis of hydrophobic OA-CFO nanoparticles

Oleic acid coated cobalt ferrite (OA-CFO) nanoparticles have been synthesized by hydrothermal method using cobalt nitrate hexahydrate ($Co(NO_3)_2 \cdot 6H_2O$) and ferric nitrate nonahydrate ($Fe(NO_3)_3.9H_2O$) as starting precursors and oleic acid as a surfactant. A solution of 1.5 mmol $Co(NO_3)_2.6H_2O$ and 3 mmol $Fe(NO_3)_3.9H_2O$ in ethanol was mixed with a 15 mmol NaOH solution and 0.2 M oleic acid. The resultant reaction solution was thoroughly stirred and poured into a Teflon lined stainless steel autoclave. The autoclave was placed into an oven for 15 hours, which was preheated at 220 $^{O}$C. After cooling the liquid phase was discarded and synthesized nanoparticles were thoroughly washed with hexane and ethanol. Each time, the synthesized CFO nanoparticles were separated from the liquid using a permanent magnet.

## 2.2 Transforming OA-CFO nanoparticles into CA-CFO nanoparticles by mechanochemical milling

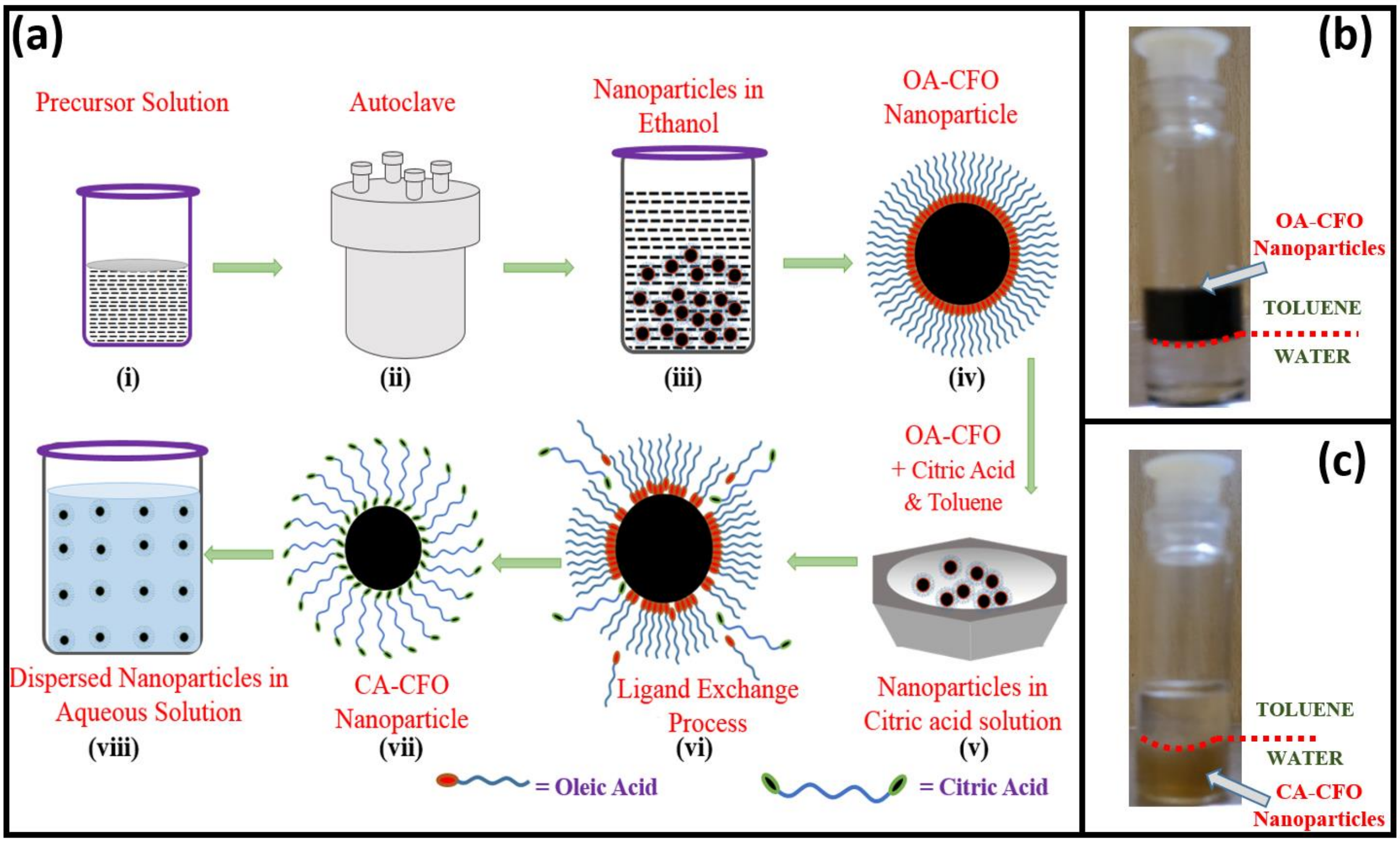

**Fig. 1** (a) Schematic of steps in the synthesis of OA-CFO and CA-CFO nanoparticles, (b) OA-CFO and (c) CA-CFO nanoparticles in toluene and water.

In order to exchange the ligand coated on the OA-CFO nanoparticles (i.e. Oleic acid) by citric acid, the OA-CFO nanoparticles were mixed with citric acid (1:5 wt/wt). Liquid assisted grinding was conducted by adding toluene to this mixture, taking the ratio of liquid volume to mass of solid (OA-CFO) nanoparticles as 5 μL $mg^{-1}$ and the mixture was grinded for 30 minutes at room temperature. The modified nanoparticles were washed with a solution of acetone and ethanol. The finally obtained nanoparticles were dispersible in water without any residue. These citric acid coated nanoparticles are named as CA-CFO in the subsequent discussion. This novel mechanochemical ligand exchange is single step and rapid compared to the conventional ligand exchange methods (Wang et al. 2014).The schematic of steps in the synthesis of OA-CFO nanoparticles and conversion of these nanoparticles into CA-CFO nanoparticles by the novel mechanochemical ligand exchange process is shown in Fig. 1 (a). The as synthesized OA-CFO nanoparticles makes stable colloidal solution when dispersed in organic nonpolar solvent, but not in water or polar solvents. The modified CA-CFO nanoparticles are well dispersible in water or in any other polar solvent but not in nonpolar solvents. Fig. 1 (b) and 1 (c) shows the macroscopic observations of OA-CFO and CA-CFO nanoparticles before and after the phase transfer. Fig. 1(b) confirms that the as synthesized OA-CFO nanoparticles are soluble in nonpolar organic solvents like toluene or oil only and are immiscible in water. The presence of OA-CFO nanoparticles can be observed by the brown/black colouring of the toluene layer over water, which verifies the hydrophobic and Oleophilic nature of the OA-CFO nanoparticles. After ligand exchange with citric acid the modified CA-CFO nanoparticles are soluble in water, but not soluble in organic solvents (Fig. 1(c)) which is an evidence of hydrophilic nature of the CA-CFO nanoparticles. The OA-CFO/CA-CFO nanoparticles does not go to the water/organic phase even after shaking that confirms the colloidal stability of these nanoparticles in nonpolar/polar solvents.

### 2.3 Preparation of Water in Oil and Oil in Water emulsions

Surfactant-stabilized water in oil (W/O) and oil in water (O/W) emulsions were prepared by taking toluene as oil phase. Toluene in water emulsion was prepared by mixing water and toluene (99:1, v/v) with the addition of 0.1 mg of SDS per ml of emulsion. Water in toluene emulsion was prepared by mixing water and toluene (1:99, v/v) with addition of 1 mg of Span-80 per ml of emulsion. The mixtures were stirred rigorously for 3 h. This gives very stable toluene in water and water in toluene emulsions which are referred as O/W and W/O emulsions in the subsequent discussion.

### 2.4 Characterization methods

The crystallographic studies of OA-CFO and CA-CFO nanoparticles have been carried out by Rigaku Ultima IV X-ray diffractometer (XRD) with CuKα radiation ($\lambda$= 1.54 Å) operated at 40 kV and 20 mA. The samples were taken in powder form and the XRD patterns were recorded in gonio scan mode for $2\theta$ value ranging from 20º to 80º with a step size of 0.05 º and scanning speed 6 º /min. The morphology of the nanoparticle samples were characterized by using a JEOL JEM-2200-FS Transmission electron microscope (TEM). Magnetic measurements of the CFO nanoparticles were performed at room temperature using the Alternating Gradient Magnetometer (AGM), in the magnetic field range of -1 to 1 Tesla. For the magnetic measurements, the CA-CFO nanoparticles were added in DI water (20 mg/mL) and OA-CFO nanoparticles were added in toluene (20 mg/mL) and then drop casted (20 µL) onto glass substrates (0.4cm × 0.4 cm) and dried properly. This step was repeated 5 times and finally obtained samples were used for magnetic measurements. FTIR studies of OA-CFO and CA-CFO nanoparticles were carried out using Thermo Scientific™ Nicolet™ iS™ 50 FT-IR Spectrometer. High performance liquid chromatography deionized water droplets (∼3 µL) with a typical resistivity value of 18.2 MΩ cm at 25 °C were deposited on the sample surfaces for contact angle measurements. The image of the droplet on sample was captured using CMOS camera equipped with a magnifying lens. The contact angle of the droplet was measured using the ImageJ software (National Institute of Health, USA), with a plugin Drop Shape Analysis.

## 3. Results and discussion

### 3.1 XRD and TEM studies

XRD patterns of OA-CFO and CA-CFO nanoparticles are shown in Fig. 2. The peaks observed at $2\theta = 30.22^{o}, 35.48^{o}, 43.18^{o}, 53.56^{o}, 57.12^{o}$ and $62.67^{o}$ corresponds to (220), (311), (400), (422), (511) and (440) planes of spinel $CoFe_2O_4$ respectively (JCPDS No. 22-1086), which confirms the formation of single phase cubic spinel structure of $CoFe_2O_4$ nanoparticles. The average crystallite size of CFO nanoparticles was determined by Scherrer's formula (Munjal and Khare 2016);

$$D_{XRD} = \frac{0.9\lambda}{\beta \cos\theta} \quad (1)$$

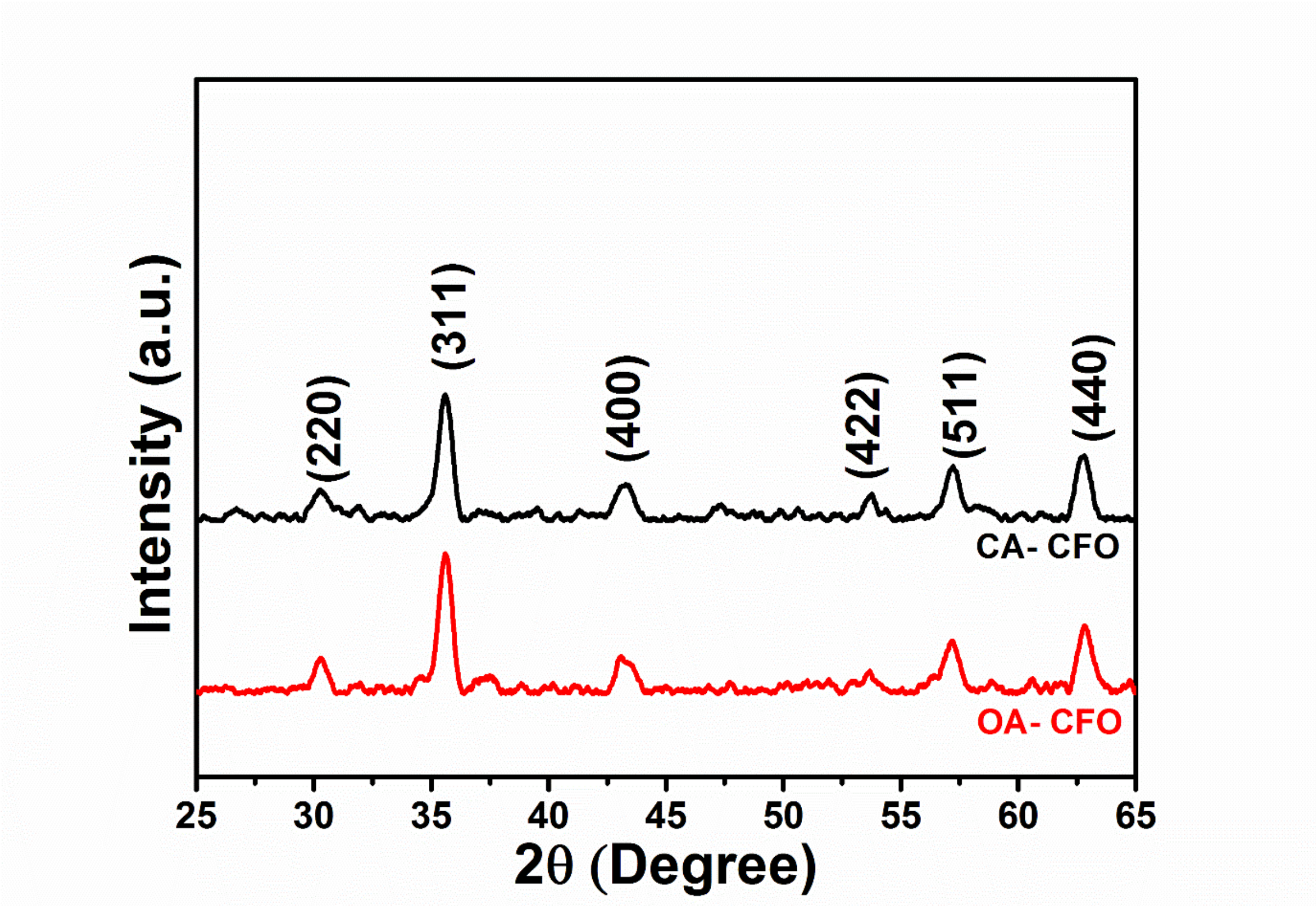

**Fig. 2** X-Ray diffraction patterns of OA-CFO and CA-CFO nanoparticles.

Where β represents the full width at half maximum of the XRD peak, λ is the wavelength of X-ray (1.542 Å), θ is the Bragg's angle, and $D_{XRD}$ is the average crystallite size.

The average crystallite sizes for OA-CFO and CA-CFO nanoparticles are found as ~10.6 nm and ~10.1 nm respectively.

. Fig. 3 shows the TEM images, histogram of particle size distribution and High Resolution TEM (HRTEM) images of OA-CFO and CA-CFO nanoparticles. Clear fringes can be seen in the HRTEM images of OA-CFO and CA-CFO nanoparticles that corresponds to (400) and (440) planes of inverse spinel cobalt ferrite structure and confirms good crystalline growth of nanoparticles.

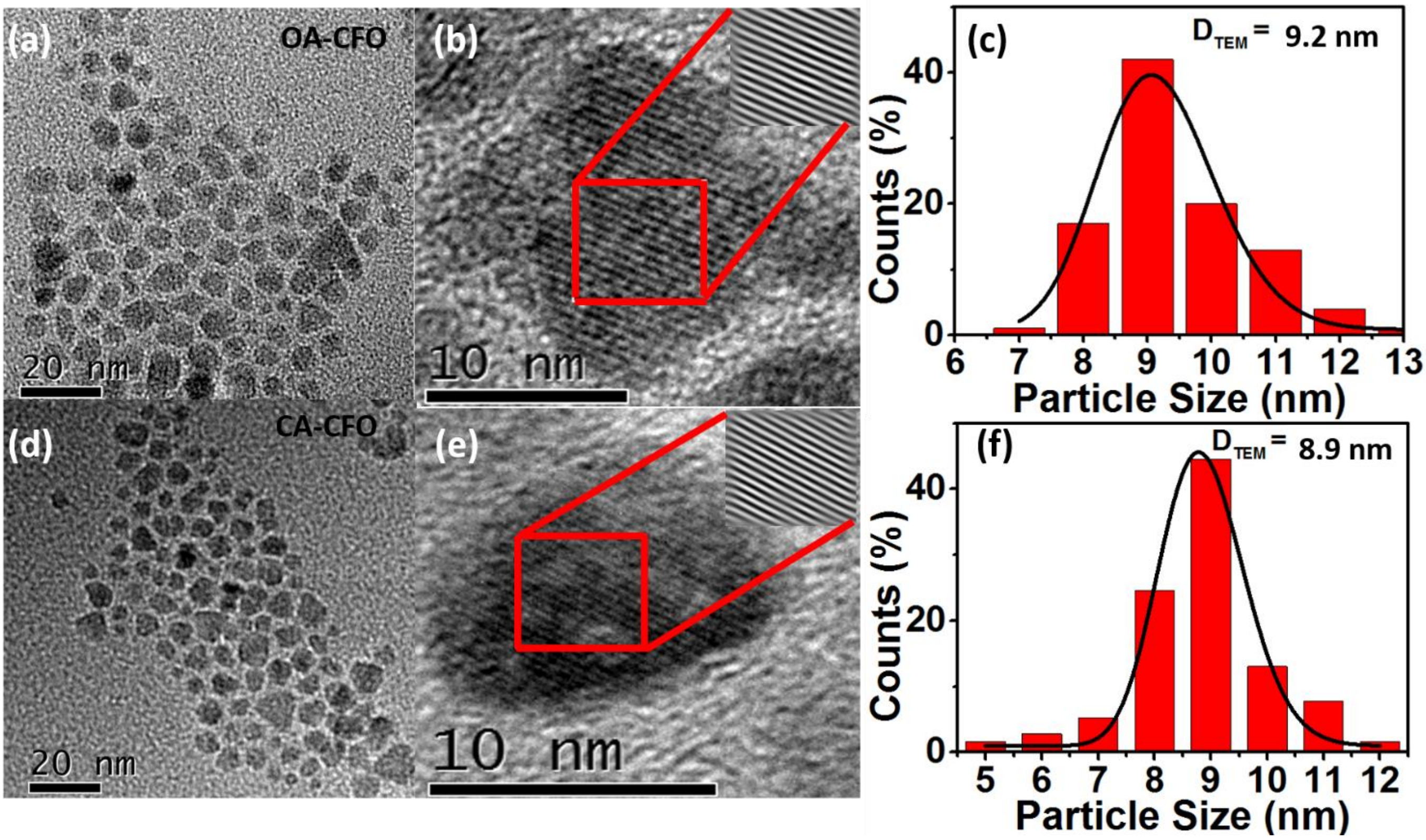

**Fig. 3** TEM images of (a) OA-CFO and (d) CA-CFO nanoparticles. High resolution TEM images of (b) OA-CFO and (e) CA-CFO nanoparticles. The inset in (b) and (e) shows the fast fourier transform (FFT) of the selected section. Fig. 3 (c) and 3 (f) shows the log normal distribution of particle size of OA-CFO and CA-CFO nanoparticles.

It is clear from the size distribution of OA-CFO and CA-CFO nanoparticles that the size of maximum number of particles lies within a very narrow range i.e. from 7 nm to 12 nm.

The particle size distribution of OA-CFO nanoparticles shows a log normal distribution peak ($D_{TEM}$) at ~ 9.2 nm whereas for CA-CFO it is at ~ 8.9 nm.

### 3.2 Magnetic Measurements

The magnetization vs. magnetic field (M-H) loops for OA-CFO and CA-CFO nanoparticles at room temperature are shown in Fig. 4. Both particles shows the ferromagnetic behaviour. The value of the $M_s$ and $H_c$ for OA-CFO nanoparticles are found as ~ 53 emu/gm and ~ 0.052 Tesla respectively, whereas for CA-CFO nanoparticles, these are ~ 49 emu/gm and ~ 0.048 Tesla respectively.

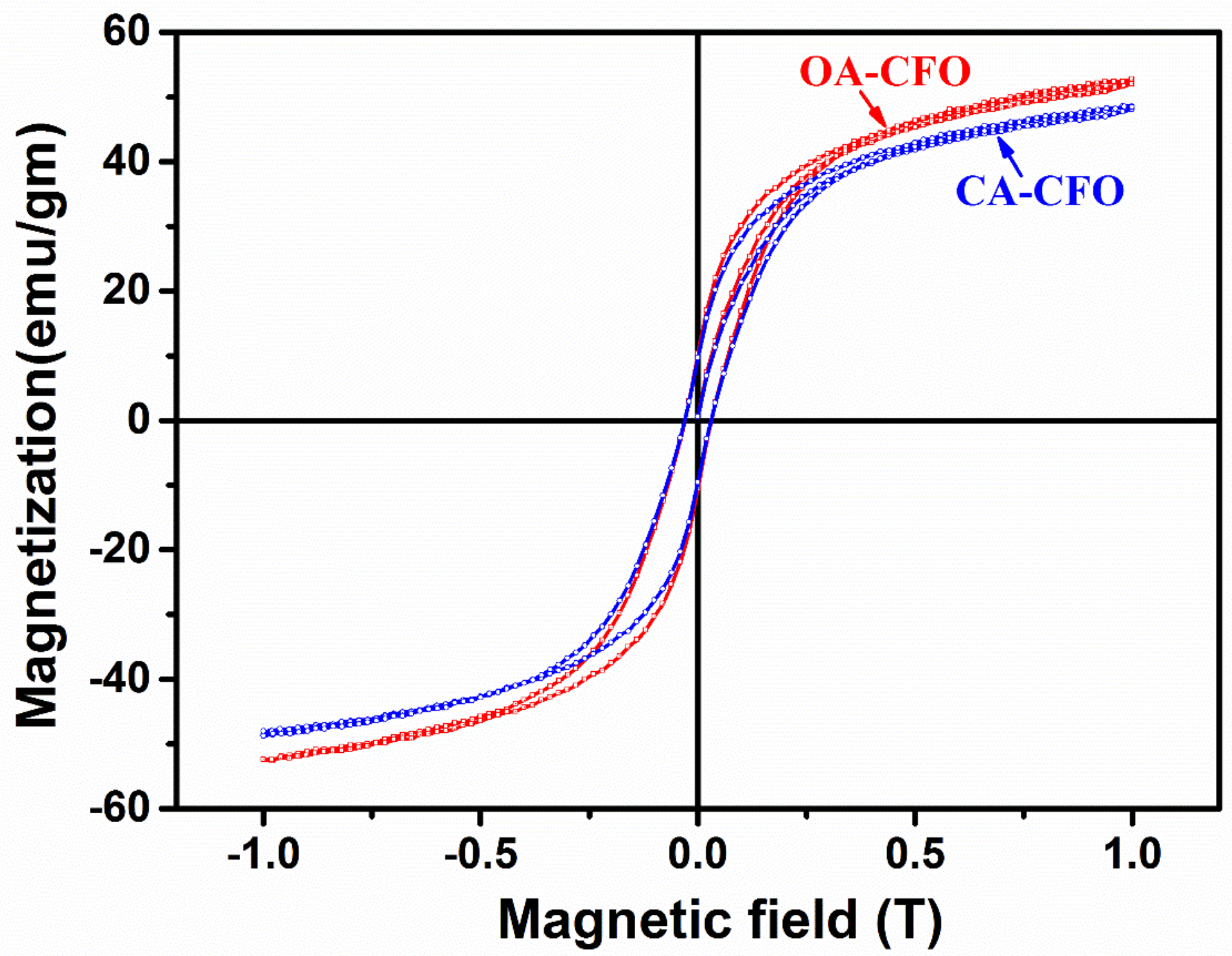

**Fig. 4** M-H loops for OA-CFO and CA-CFO nanoparticles.

A decrease in magnetization of CFO nanoparticles after the ligand exchange completion was observed. This reduction in saturation magnetization from OA-CFO (53 emu/gm) to CA-CFO (49 emu/gm) is attributed to the mechanochemical ligand exchange of the OA-CFO nanoparticles. During ligand exchange process removal of some surface cations takes place and formation of a new Ligand-metal bond may further contribute in the decrease of the Ms values (Palma et al. 2015). The ligand exchange process reduces the core size (~ 4 %) of the nanoparticle which was confirmed by the TEM images.

### 3.3 FTIR Studies

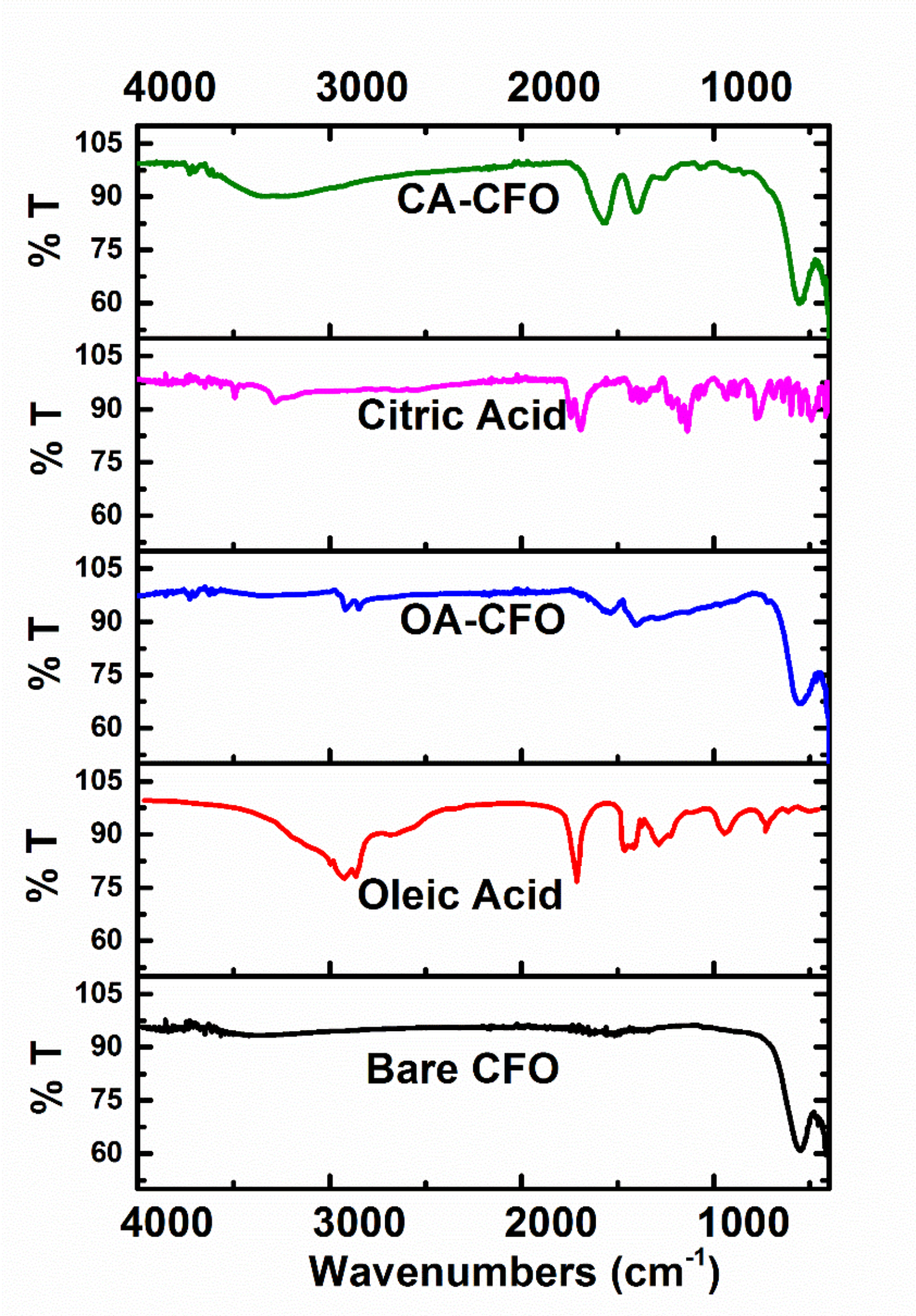

**Fig. 5** FTIR spectra of bare CFO nanoparticles, neat oleic acid, OA-CFO nanoparticles, neat citric acid and CA-CFO nanoparticles.

Fig. 5 shows the FTIR spectra of bare CFO nanoparticle (synthesized by the same route without OA), neat oleic acid, OA-CFO nanoparticles, neat citric acid and CA-CFO nanoparticles in the wavenumber ranging from 4000 $cm^{-1}$ to 500 $cm^{-1}$.

The FTIR spectra of neat oleic shows distinct peaks at 2850 $cm^{-1}$ and 2920 $cm^{-1}$ (due to stretching of =CH- alkene group ) and broad feature from 3500 $cm^{-1}$ to 2500 $cm^{-1}$ is characteristic of the O–H stretching band of the acid and the broadness of this band is caused by intramolecular hydrogen bonding . The characteristic carbonyl band appears at 1713 $cm^{-1}$ . The two absorption peaks that appear at 1418 $cm^{-1}$ and 1285 $cm^{-1}$ are due to O–H bending and C–O stretching (Zhang, He, and Gu 2006). For bare CFO nanoparticles, the only one peak is present at 590 $cm^{-1}$ and can be attributed to stretching of metal ion-oxygen bond (Pilapong et al. 2015).

The presence of oleic acid on as synthesized oleic acid coated samples OA-CFO nanoparticles was confirmed by two $-CH_2$ stretching near 2920 $cm^{-1}$ and 2850 $cm^{-1}$ present in FTIR spectra of the OA-CFO sample (Wu et al. 2004). The carboxyl group has characteristic bands of the asymmetric stretch ($\upsilon_{as[-COO^-]}$) and the symmetric stretch ($\upsilon_{s[-COO^-]}$). Two bands appears near 1538 and 1410 $cm^{-1}$, corresponds to $\upsilon_{as[-COO^-]}$ and $\upsilon_{s[-COO^-]}$. $-COO^-$ and $-CH_3$ stretching bands confirms the presence of OA on coated samples (Patil et al. 2014). The peak near 1710 $cm^{-1}$ represents the C=O stretch band of the carboxyl group present in oleic acid, which was found absent in the spectrum of the oleic acid coated nanoparticles. This indicates that oleic acid was chemisorbed onto the surface of CFO nanoparticles via its carboxylate group (Zhang et al. 2006).

Now, during the mechanochemical milling, as the ligand exchange process occurs the citric acid starts replacing the oleic acid from the surface of CFO nanoparticles. The citric acid is a tridentate ligand that has three - $COO^-$ groups on a relatively small carbon chain (Tang al. 2013). After surface modification by citric acid three absorption peaks were observed near 3727, 1576, and 1402 $cm^{-1}$. The former is attributed to the stretching band of hydroxyl group (-OH) and the others were attributed to the $\upsilon_{as[-COO^-]}$ and $\upsilon_{s[-COO^-]}$ stretching band of the carboxyl group, respectively (Liao et al. 2015). This indicates that the surface of the CA-CFO nanoparticles was covered with carboxylate species of citric acid. An intense peak at ~590 $cm^{-1}$ is observed, that can be attributed to the stretching of the metal ion at the tetrahedral A-site, $M_A \leftrightarrow O$ (Pilapong et al. 2015).

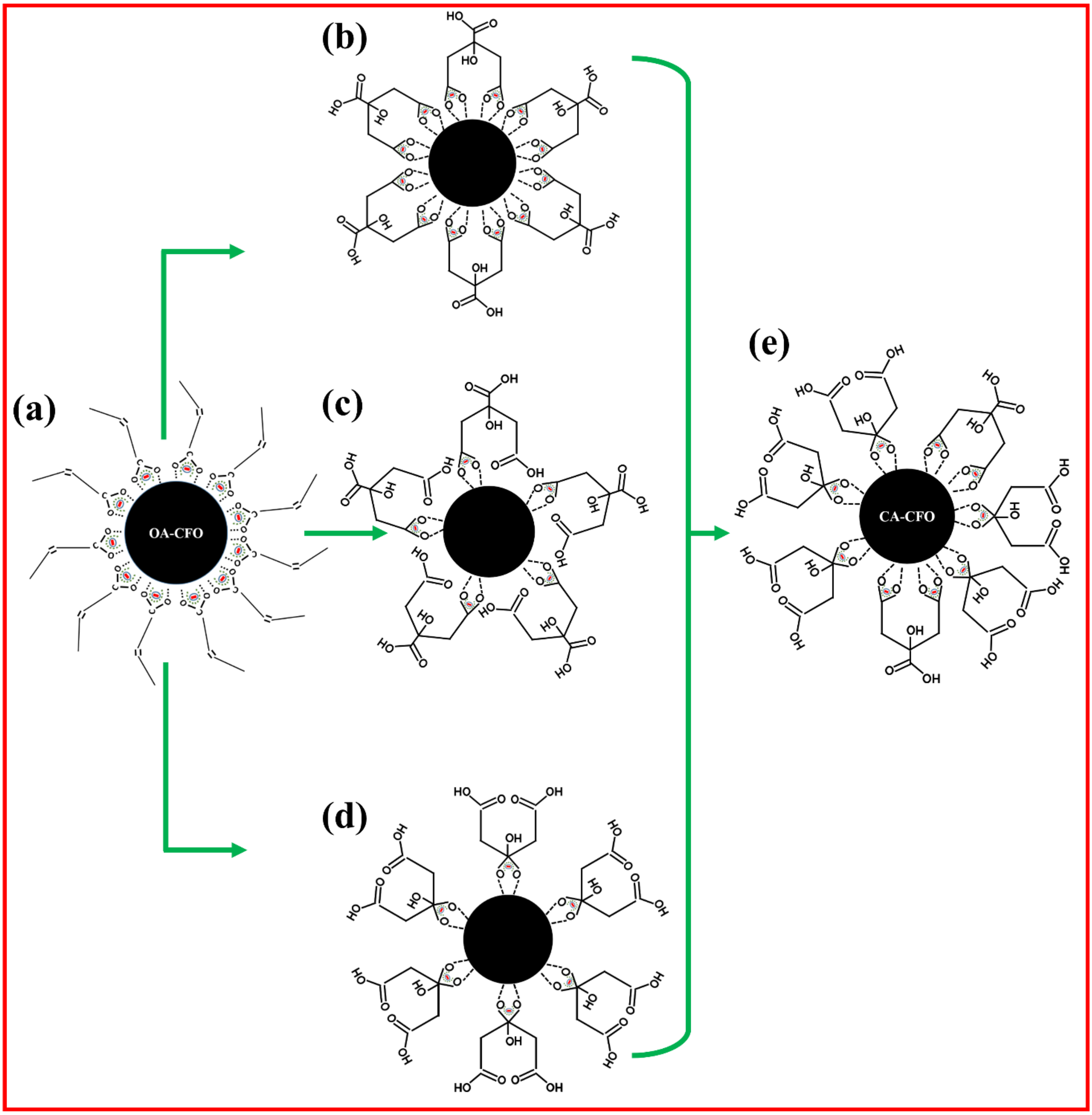

**Fig. 6** Representation of different possibilities of attachment of ligands on the surface in the case of (a) OA-CFO nanoparticles and (b)-(e) CA-CFO nanoparticles.

It is also evident from the FTIR spectra that, after the surface modification, two $-CH_3$ stretching near 2920 $cm^{-1}$ and 2850$cm^{-1}$ becomes weaker (Palma et al. 2015) , which confirms that ligand exchange takes place between the oleic acid capping and citric acid and after this exchange only citric acid layer covers the nanoparticles.

There are many possible ways of attachment of citric acid on the surface of cobalt ferrite nanoparticles. The citric acid can get attached with the nanoparticle by one of the $COO^-$ groups present at the corner or middle of the chain as shown in Fig. 6 (c) and 6 (d), or by two $-COO^-$ groups present at the corner of the carbon chain [Fig. 6 (b)]. However all these conditions or a mixed condition [Fig. 6(e)] are possible and in each case at least one $-COO^-$ group remains on the outer surface of the CA-CFO nanoparticles.

The hexane dispersion of OA-CFO nanoparticles and water dispersion of CA-CFO nanoparticles remains stable even after 6 months. This stability was confirmed by zeta potential ($\zeta$) measurements. The zeta potential ($\zeta$) values of as synthesized OA-CFO sample in hexane was − 38 mV and after 6 months its value was obtained as − 36 mV. It is clear that $\zeta$ values are very large even after a long time that gives the stability to OA-CFO nanoparticles in organic solvents like hexane. On the other hand, the $\zeta$ of CA-CFO nanoparticle changes from − 39 mV to − 27 mV after 6 months but still remains sufficiently large. The large enough negative value of $\zeta$ even after a long time confirms the stability of the CA-CFO nanoparticles in water.

Hydrodynamic diameter ($D_H$) and histograms for grain size distribution (as synthesized and after six months) for CA-CFO nanoparticles suspended in hexane and CA-CFO nanoparticles suspended in water are shown in the Fig. 7.

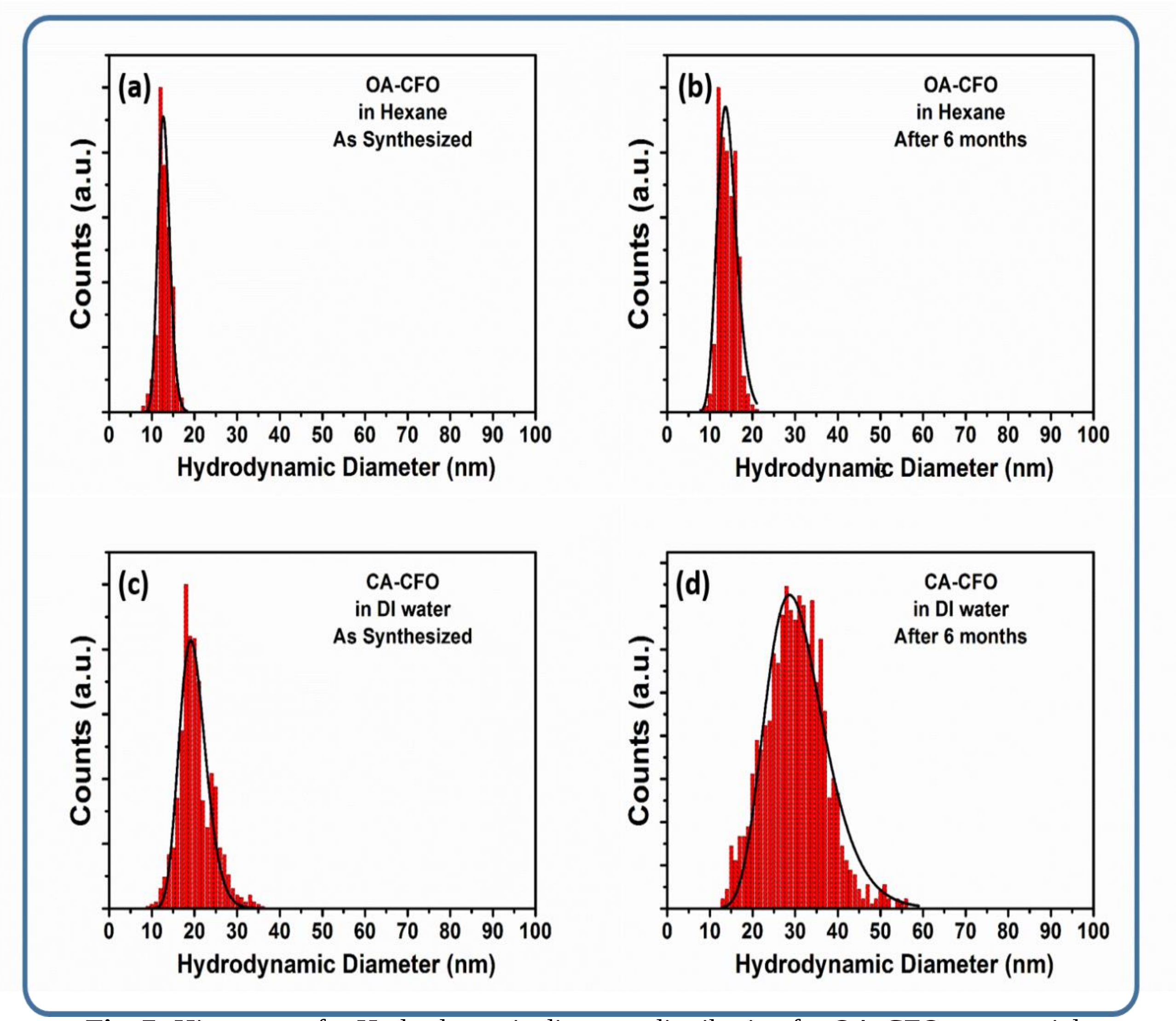

**Fig. 7** Histograms for Hydrodynamic diameter distribution for OA-CFO nanoparticles suspended in hexane (a) as synthesized, (b) after 6 months and for CA-CFO nanoparticles suspended in water (c) as synthesized, (d) after 6 months.

It is to note that hydrodynamic diameter of OA-CFO in hexane is larger than that of CA-CFO nanoparticles in water.

This can be explained by the nature of bonding between ligand present on the surface of nanoparticles and the solvent molecules. As hexane is an organic/nonpolar solvent and thus mixing of the OA-CFO nanoparticles in it that has free "-$CH_3$" groups on their outer surface, it does not make any hydrogen bond with the surface of the nanoparticle. On the other hand, negatively charged "-$COO^-$" group present on the outer surface of the CA-CFO nanoparticle makes hydrogen bonds with the water solvent that increases the $D_H$ values of CA-CFO nanoparticles in water medium compared to OA-CFO nanoparticles in hexane.

It is also observed that hydrodynamic diameters of OA-CFO and CA-CFO nanoparticles are larger than the diameters of the nanoparticles obtained by the TEM images. As the hydrodynamic diameters gives us information of the CFO core along with coating material and the solvent layer attached to the particle. While estimating size by TEM, this hydration layer is not present, hence we get information only about the CFO core.

### 3.4 Contact angle measurements

Contact angle ($\theta_C$) measurements for OA-CFO and CA-CFO have been performed for studying the hydrophobic or the hydrophilic behaviour of any surface. For $\theta_C$ measurements, the thin films of OA-CFO and CA-CFO nanoparticles were deposited by mixing the nanoparticles in isopropanol. For this, 100 mg of OA-CFO or CA-CFO nanoparticles were mixed in 20 ml isopropanol and then spray coated on a glass substrate kept at 100 ºC. A water droplet of 3 µL was deposited on these films for studying the hydrophobic/hydrophilic behaviour. Fig. 8 shows the photographs of water droplets on the OA-CFO and CA-CFO sample surface for different duration of time after the water droplet was left on the surface. Fig. 9 shows the measured contact angles ($\theta_C$) for the water droplet on the two surfaces for different deposition time.

The water contact angle on the OA-CFO sample's surface was very high (~145$^o$) and water droplet deposited on the surface was almost perfect sphere, that confirms the hydrophobic behaviour of the OA-CFO sample (Sharp et al. 2014). This hydrophobic nature of these OA-CFO nanoparticles can be attributed to the presence of $-CH_3$ on the outer surface of these nanoparticles. One of the two ends of OA has $-COO^-$ group whereas the other end has $-CH_3$ group. The $-COO^-$ group behaves like a hydrophilic group whereas the $-CH_3$ group behave like a hydrophobic group (Lei et al. 2015). In the case of OA-CFO the – $COO^-$ groups are the anchoring group, bonded to the surface of CFO nanoparticles and $–CH_3$ groups remain free on the outer surface of the OA-CFO nanoparticles. The presence of hydrophobic groups at the outer layer of the OA-CFO nanoparticles makes them hydrophobic and Oleophilic.

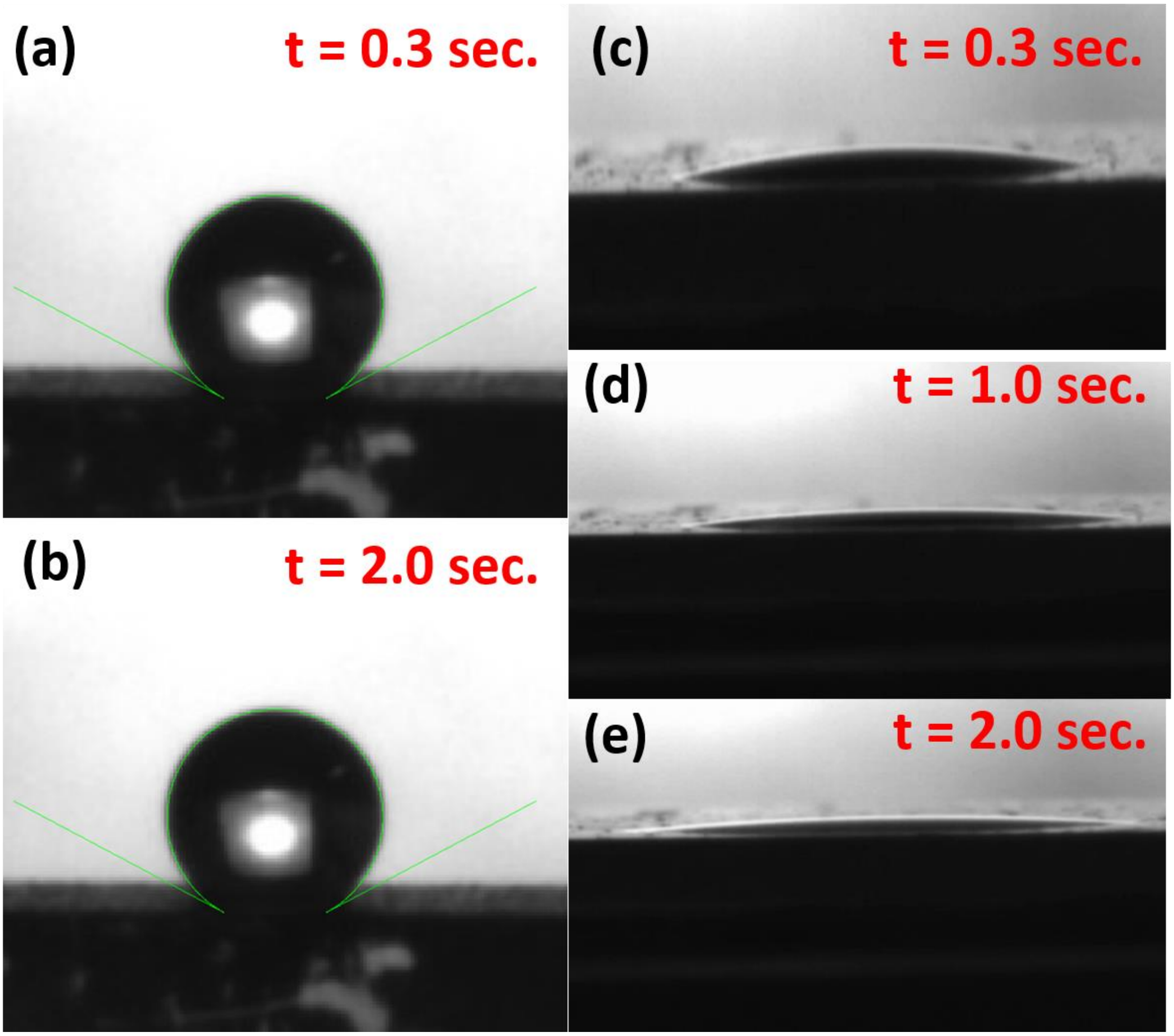

**Fig. 8** The photographs of the 3 µL water droplet on the film of OA-CFO nanoparticles at (a) 0.3 sec. and (b) 2.0 sec. and on the film of CA-CFO nanoparticles at (c) 0.3 sec. (d) 1.0 sec. and (e) 2.0 sec., after water droplet left on the surface.

For the CA-CFO nanoparticles surface the water contact angle reduces to $8^o$ in a few seconds, which confirms the hydrophilic behaviour of CA-CFO nanoparticles (Sharp et al. 2014). This hydrophilic behaviour can be attributed to the presence of – $COO^-$ groups. After the mechanochemical ligand exchange a significant change in the surface chemistry of the nanoparticles and hence in the water contact angle with the CA-CFO have been observed.

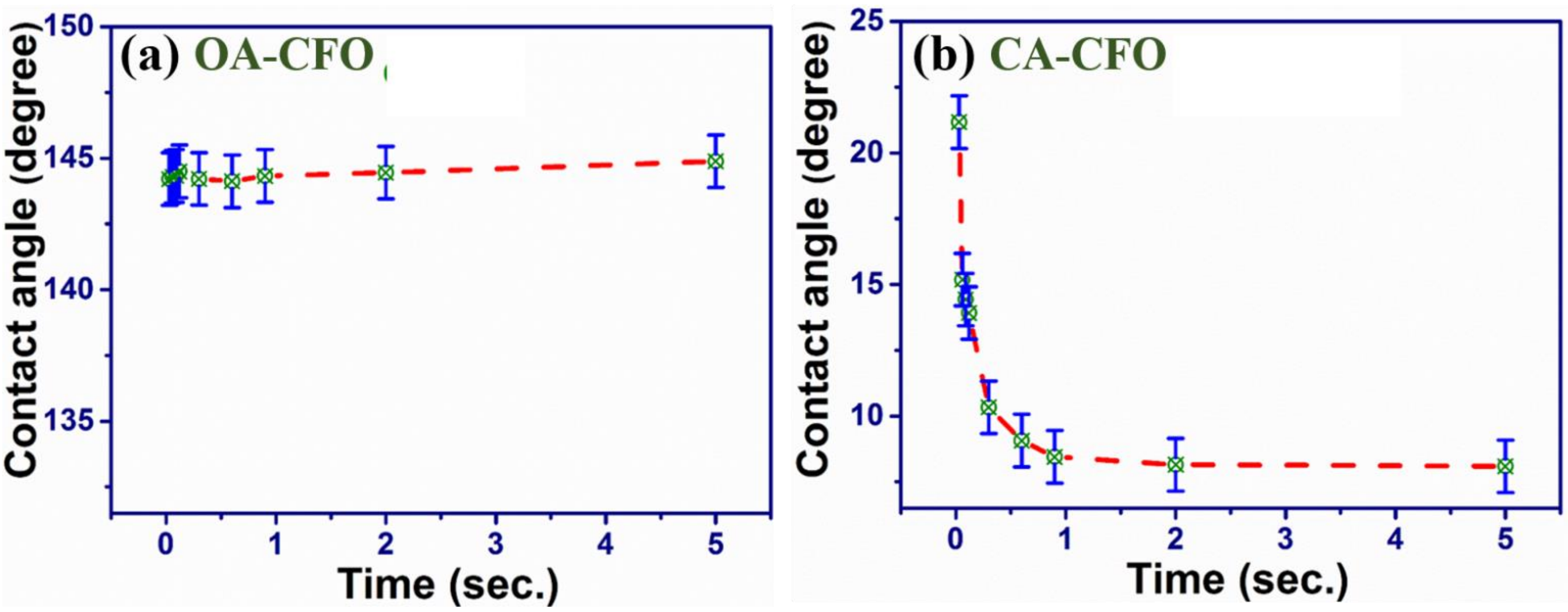

**Fig. 9** Variations in water contact angle with time for the thin film of spray coated nanoparticles of (a) OA-CFO and (b) CA-CFO.

As clear by the FTIR studies discussed above at least one -$COO^-$ group per citric acid molecule remains on the outer surface of the CA-CFO nanoparticles, that makes the outer surface of the CA-CFO nanoparticle negatively charged and the CA-CFO nanoparticles show hydrophilic behaviour.

### 3.5 Demulsification Tests and Recycling

The potentiality of OA-CFO and CA-CFO nanoparticles in demulsification of O/W and W/O emulsions was examined by placing these nanoparticles in emulsion. In order to check the stability of O/W (oil in water) and W/O (water in oil) emulsions we prepared fresh samples of the two emulsions by the method as described in experimental section. Fig. 10 (a) shows the fresh samples of O/W and W/O and Fig. 10 (b) shows the same emulsions after 10 h of the preparation. It is clear from the figures that the emulsions are stable with time without demulsification even after 10 h. Fig. 10 (c) (1) and 10 (d) (1) shows the photographs of oil in water and water in oil emulsions. For testing the demulsification potential of these coated CFO (OA-CFO and CA-CFO) nanoparticles, the nanoparticles were mixed in O/W and W/O emulsions (40 mg CFO nanoparticles for per ml of emulsion) respectively and ultrasonicated for 60 minutes (Fig. 10 (c) (2) and Fig. 10 (d) (2)). After adding the nanoparticles and ultrasonication, the mixture gives a uniform colour, indicating that the nanoparticles were well dispersed in the emulsion.

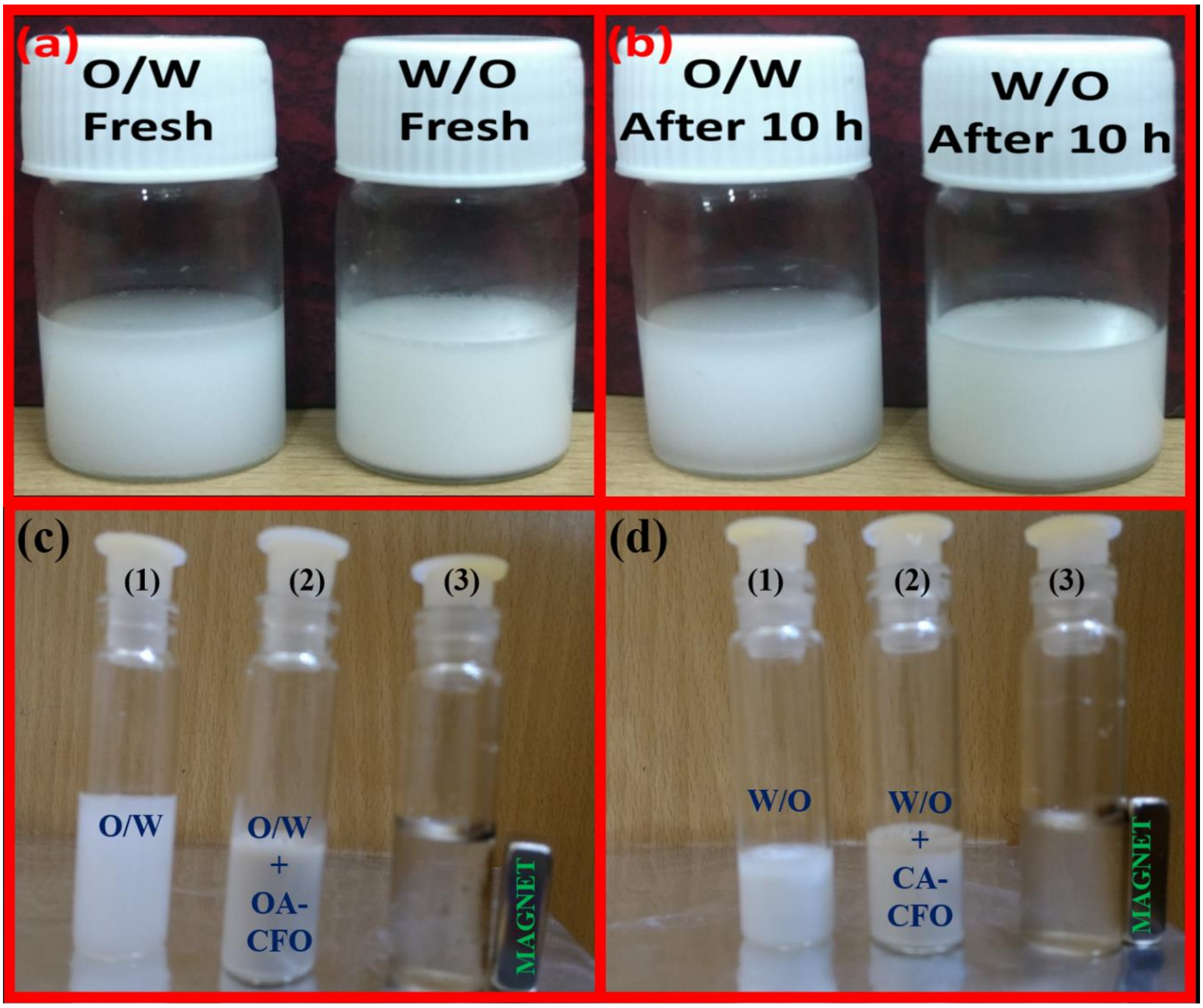

**Fig. 10** Stability test of the O/W and W/O emulsions. (a) fresh samples and (b) samples after 10 h. Photographs of (c) (1) O/W emulsion, (2) OA-CFO dispersed in O/W emulsion

and, (3) after demulsification of O/W emulsions. (d) (1) W/O emulsion, (2) CA-CFO dispersed in W/O emulsion and, (3) after demulsification of W/O emulsions.

As OA-CFO nanoparticles are Oleophilic in nature, they bind small oil droplets with them in O/W emulsion. Similarly CA-CFO nanoparticles bind water droplets with them in W/O emulsion due to their hydrophilic nature. When a hand magnet was placed near the mixture solution, the black OA-CFO and CA-CFO nanoparticles together with their tagged water/oil phase were attracted towards the magnet and the system became nearly transparent and colourless (Fig. 10 (c) (3) and (d) (3). This demonstrates that the toluene droplets in O/W and water droplets in W/O emulsions were very effectively separated from the emulsion using OA-CFO and CA-CFO nanoparticles respectively. After the demulsification, the magnetic nanoparticles were collected using a small permanent magnet and washed using chloroform ($CHCl_3$) and ethanol for 3 times. Before reusing the nanoparticles for the demulsification, FTIR studies were carried out. No significant change in the surface properties of the nanoparticles was observed and the nanoparticles have been reused for the demulsification.

## 4. Conclusion

We have successfully synthesized oleic acid capped, single phase, uniform size (~9 nm) CFO nanoparticles by hydrothermal method and demonstrated the conversion of these nanoparticles to a citric acid (CA) coated CFO nanoparticles in short time (~30 min.) using a rapid and novel mechanochemical ligand exchange method. The FTIR spectra confirms the presence oleic acid on OA-CFO nanoparticles and exchange of oleic acid with citric acid. The attachment of anchoring groups on OA-CFO and CA-CFO nanoparticles is also explained. Contact angle measurements confirm the hydrophobic and hydrophilic behaviour of oleic acid coated and citric acid coated CFO nanoparticles respectively. It is demonstrated that the hydrophobic and hydrophilic behaviour of these magnetic nanoparticles can be used for the demulsification of oil in water and water in oil emulsions. These nanoparticles can be reused after washing them and no significant change in their surface properties was observed after the demulsification. FTIR studies revealed that for OA-CFO and CA-CFO nanoparticles, the $-CH_3$ and $-COO^-$ groups are exposed to the surrounding solvent that makes them hydrophobic (oleophilic) and hydrophilic respectively. The novel mechanochemical milling method is proposed as a rapid and efficient mthod for ligand exchange.

## 5. Acknowledgments

The authors are thankful to the DeitY (project no. RP02395) and one of the authors (Sandeep Munjal) is thankful to Council of Scientific and Industrial Research (CSIR), New Delhi for Senior Research Fellowship (SRF) Grant (09/086(1179)/2013-EMR1 ). It is declared that authors have no other conflict of interest.

## Figure Captions

**Fig. 1** (a) Schematic of steps in the synthesis of OA-CFO and CA-CFO nanoparticles, (b) OA-CFO and (c) CA-CFO nanoparticles in toluene and water.

**Fig. 2** X-Ray diffraction patterns of OA-CFO and CA-CFO nanoparticles.

**Fig. 3** TEM images of (a) OA-CFO and (d) CA-CFO nanoparticles. High resolution TEM images of (b) OA-CFO and (e) CA-CFO nanoparticles. The inset in (b) and (e) shows the fast fourier transform (FFT) of the selected section. Fig. 3 (c) and 3 (f) shows the log normal distribution of particle size of OA-CFO and CA-CFO nanoparticles.

**Fig. 4** M-H loops for OA-CFO and CA-CFO nanoparticles.

**Fig. 5**. FTIR spectra of bare CFO nanoparticles, neat oleic acid, OA-CFO nanoparticles, neat citric acid and CA-CFO nanoparticles.

**Fig. 6** Representation of different possibilities of attachment of ligands on the surface in the case of (a) OA-CFO nanoparticles and (b)-(e) CA-CFO nanoparticles.

**Fig. 7** Histograms for Hydrodynamic diameter distribution for OA-CFO nanoparticles suspended in hexane (a) as synthesized, (b) after 6 months and for CA-CFO nanoparticles suspended in water (c) as synthesized, (d) after 6 months.

**Fig. 8**. The photographs of the 3 μL water droplet on the film of OA-CFO nanoparticles at (a) 0.3 sec. and (b) 2.0 sec. and on the film of CA-CFO nanoparticles at (c) 0.3 sec. (d) 1.0 sec. and (e) 2.0 sec., after water droplet left on the surface.

**Fig. 9** Variations in water contact angle with time for the thin film of spray coated nanoparticles of (a) OA-CFO and (b) CA-CFO.

**Fig. 10** Stability test of the O/W and W/O emulsions. (a) fresh samples and (b) samples after 10 h. Photographs of (c) (1) O/W emulsion, (2) OA-CFO dispersed in O/W emulsion and, (3) after demulsification of O/W emulsions. (d) (1) W/O emulsion, (2) CA-CFO dispersed in W/O emulsion and, (3) after demulsification of W/O emulsions.